# Oscillation quenching in third order Phase Locked Loop coupled by mean field diffusive coupling

S Chakraborty[1*], M Dandapathak[2] and B C Sarkar [3]
[1]Physics Department, Bidhan Chandra College, Asansol- 713304, WB, INDIA
[2]Physics Department, Hooghly Mohsin College, Chinsurah, Hooghly-712101, WB, INDIA.
[3]Physics Department, Burdwan University, Burdwan-713104, WB, INDIA
* e-mail: saumenbcc@gmail.com

**Abstract:** We explore analytically the oscillation quenching phenomena (amplitude death and oscillation death) in a coupled third order phase locked loop (PLL) both in periodic and chaotic mode. The phase locked loops are coupled through mean field diffusive coupling. The lower and upper limits of the quenched state are identified in the parameter space of the coupled PLL using Routh-Hurwitz technique. We further observe that the ability of convergence to the quenched state of coupled PLLs depends on the design parameters. For identical system both the system converges to homogeneous steady state whereas for non-identical parameter values they converge to inhomogeneous steady state. It is also observed that for identical systems the quenched state is wider than non-identical case. When the systems parameters are so chosen that each isolated loops are chaotic in nature, in that case we observe the quenched state is relatively narrow. All these phenomena are also demonstrated through numerical simulations.

**Lead Paragraph:**
Oscillation quenching is one of the important emergent phenomena of coupled dynamical systems. Therefore, the study on the behavior of different coupled system widely increased in recent. Here we explore the oscillation quenching phenomena in a coupled third order phase locked loop having resonant type low pass filter. Mean-field diffusive coupling scheme has been used for coupled them. The oscillation quenched states of different types (oscillation death, amplitude death) are obtained for identical and non-identical coupled system. The range of quenched states of the system has been estimated in terms of coupling parameter with different design parameter values. The dynamics of such coupled third order phase locked loop has been studied analytically and numerically in the parameter space of the system. Both the obtained results are in agreement with each other. Further, we study the dynamics of the coupled system when they are chaotic in nature.

1. **Introduction:**

The study on the behavior of different coupled system widely increased in the recent past. Coupled dynamical systems show a wide variety of complex collective behaviors like synchronization, phase-locking, oscillation quenching, etc. Such co-operative phenomena of coupled systems are of significant interest in the field of physics, biology and engineering applications [1-9]. Studies on coupled system are also important due to the fact that most of the natural systems are non-isolated.

Oscillation quenching is one of the important emergent phenomena in which cessation of oscillation to the steady state occur under some proper parametric conditions of coupled dynamical systems. Such oscillation suppression is relevant in neuronal disease and also in stabilization of system dynamics [1, 2]. So, it has been characterized and extensively studied by the academician and the engineers of various fields.

Oscillation quenching is of two types: amplitude death (AD) and oscillation death (OD). In AD coupled systems arrive at a common stable steady state which was unstable otherwise for certain value of coupling parameter and thus form a stable homogeneous steady state (HSS). But, in the case of OD, systems populate different coupling dependent steady states and thus give rise to stable inhomogeneous steady states (IHSS) [10]. An extensive research work has been reported on oscillation quenching. But Koseska et al recently in their works [1] gave an elaborate view on these two distinct features (i.e. AD and OD).

Phase locked loops (PLLs) in analog as well as digital communication systems are of enormous importance [11, 12]. The coupled PLLs are able to overcome the limitations of coupled oscillators like intrinsic small locking bandwidth, amplitude fluctuations etc. [13]. Therefore, studies on the dynamics of coupled PLLs are also important. Several works have been carried out to study the collective behavior like synchronization, phase-locking etc. of coupled PLLs [14-18]. But the oscillation quenching phenomena has yet not been studied. PLLs with first order loop filters are widely studied and employed in practical systems, because of its unconditional stability and easy designing in practice [11, 12]. However, in case of high speed data recovery the third order PLLs (TOPLL i.e. PLLs with second order loop filters) are preferable. Such higher order PLLs are conditionally stable. Due to the variation of control parameter such PLLs show limit cycle oscillations of period-1 as well as multi period oscillations beyond the stable range and ultimately becomes chaotic [19-22].

In the present paper, we examine the dynamics of a coupled TOPLL. Here each TOPLL comprises with a resonant low pass filter (LPF). Our aim is to study oscillation quenching phenomena and define its range in different parameter space of coupled TOPLL. So far our knowledge, no works on the oscillation quenching in a TOPLL is reported up to date. To induce oscillation quenching in coupled oscillators, several coupling schemes have been proposed [2, 23]. Here we consider mean field diffusive (MFD) coupling. MFD coupling is one of the most widely used coupling method because of its presence in many natural phenomena in the field of biology, physics, and engineering [23-27]. MFD coupling is also able to remove the constraint of having parameter mismatch or time-delay coupling to obtain AD [23]. It is also noted that, the dynamical behavior of the coupled system depends both on the coupling parameters and on the design parameters of the oscillating system.

The paper has been organized in the following way. In Section 2, the differential equations describing the system dynamics for a coupled PLL system have been formulated. The system equations are written in terms of a new set of state variables which could be directly measured in experiment (as mentioned in ref. [20]). The occurrences of OD and AD in the system have been discussed in section 3. The lower and upper limits of the oscillation quenched state in parameter space of the coupled PLL have also been calculated analytically using Routh-Hurwitz technique in this section. The results of numerical simulation on the oscillation quenching of the coupled PLL system has been given in Section 4. The findings regarding OD and AD in coupling parameter space, effects of design parameters of PLL etc. have also been discussed at length here. Some concluding remarks are given in the Section 5.

   2. **Formulation of system equations:**

The authors, in ref. [20] already reported that the dynamics of an isolated TOPLL comprises with a resonant type LPF can be described by following three first order autonomous equations,

$$\frac{dx_i}{d\tau} = \Omega_{ni} - k_{ni}z_i \qquad (1a)$$

$$\frac{dy_i}{d\tau} = \sin(x_i) + (g_i - 2)y_i - (\frac{g_i-1}{g_i})z_i \qquad (1b)$$

$$\frac{dz_i}{d\tau} = g_i y_i - z_i \qquad (1c)$$

Here, the system equations are written in terms of $x_i$, $y_i$ and $z_i$ (shown in Fig-1) as three state variables. We take them as state variables since these quantities could be experimentally measured in a PLL circuit. Therefore, the prediction regarding the system dynamics could be easily verified experimentally. $x_i$, $y_i$ and $z_i$ are the phase error, voltages at the input capacitor C and at the output terminal of the resonant LPF, respectively. $z_i$ is used as the control voltage of the loop voltage control oscillator (VCO). Here we take $\tau$, $\Omega_{ni}$ and $k_{ni}$ as normalized parameters for time, frequency offset and loop gain of the PLL respectively. The normalization has been done in terms of the time constant $T(=RC)$ of the loop filter.

We can consider N numbers of such TOPLLs are interacting through mean-field diffusive coupling; the mathematical model of the coupled system is given as,

$$\frac{dx_i}{d\tau} = \Omega_{ni} - k_{ni}(z_i + d(m\bar{z} - z_i)) \qquad (2a)$$

$$\frac{dy_i}{d\tau} = \sin(x_i) + (g_i - 2)y_i - (\frac{g_i-1}{g_i})z_i \qquad (2b)$$

$$\frac{dz_i}{d\tau} = g_i y_i - z_i \qquad (2c)$$

With $i = 1,2,\ldots\ldots N$; $\bar{z} = \frac{1}{N}\sum_{i=1}^{N} z_i$ be the mean field of the coupled system. We define normalized variables $\Omega_{ni}$, $k_{ni}$ as $\Omega_i T$, $k_i T$ respectively. Here, $k_{ni}$ is the loop gain parameter taking care of VCO sensitivity, signal amplitude etc, $\Omega_{ni}$ is frequency offset of each PLL defined in terms of the input reference frequency and VCO free running frequencies and $g_i$ is the filter gain parameters of PLL. The diffusive coupling strength is given by $d$ and $m$ is the strength of mean-field. Here, $0 \le m \le 1$. We take two TOPLLs as the limiting case to study the dynamics, i.e., $N = 2$. The isolated PLL shows oscillations of multi periods for proper choice of design parameters beyond stable range.

From equation (2) we can see that the system has the following steady state value, $(x_1^*, y_1^*, z_1^*, x_2^*, y_2^*, z_2^*)$ where,

$$x_1^* = sin^{-1}(\frac{z_1^*}{g_1}), \quad y_1^* = \frac{z_1^*}{g_1}, \quad z_1^* = \frac{\beta_1\mu_1 - \beta_2\mu_2}{\mu_1^2 - \mu_2^2} \qquad (3a)$$

$$x_2^* = sin^{-1}(\frac{z_2^*}{g_2}), \quad y_2^* = \frac{z_2^*}{g_2}, \quad z_2^* = \frac{\beta_2\mu_1 - \beta_1\mu_2}{\mu_1^2 - \mu_2^2} \qquad (3b)$$

With, $\beta_1 = \frac{\Omega_{n1}}{k_{n1}}$, $\beta_2 = \frac{\Omega_{n2}}{k_{n2}}$, $\mu_1 = (1 - d + \frac{md}{2})$ and $\mu_2 = (\frac{md}{2})$.

Note that the steady state values depend on the coupling parameters, as well as on the design parameters of the PLL.

### 3. Assessment of Oscillation Quenched State:

In this section we calculated analytically the lower and upper limit of the oscillation quenched state using Routh-Hurwitz technique and predict the zone of AD/OD in the coupling parameter space.

### a. Assessment of AD:

In case of AD, for the HSS we have $x_1^* = x_2^*$, $y_1^* = y_2^*$, $z_1^* = z_2^*$. Observing the relations given in equations (3a) and (3b), it is cleared that the AD is obtained only when $g_1 = g_2$ and $\beta_1 = \beta_2$. This means the two systems are identical.

To find the stability of the HSS, we construct the linear transformation Jacobian matrix of the system at the steady state points using the relations $z_1^* = z_2^*$. The characteristic equations of the system at the above mentioned steady state values are,

$$\lambda_{1,2,3}^3 + (3 - g_1)\lambda_{1,2,3}^2 + \lambda_{1,2,3} + k_{n1}g_1(1 - d + md)\cos(x_1^*) = 0 \tag{4}$$

We apply the Routh-Hurwitz array method to the above equations. This gives the lower limit of $d$ for AD state in terms of $\Omega_n$, $k_n$, $g$ and $m$ and is given by,

$$(3 - g_1) - k_{n1}g_1(1 - d + md)\cos(x_1^*) > 0 \tag{5}$$

Replacing the term $\cos(x_1^*)$ by, $\sqrt{(1 - (\frac{z_1^*}{g_1})^2)}$ and $z_1^* = \frac{\beta_1\mu_1 - \beta_2\mu_2}{\mu_1^2 - \mu_2^2}$; we get the following condition,

$$[\frac{\beta_1\mu_1 - \beta_2\mu_2}{g_1(\mu_1^2 - \mu_2^2)}]^2 > 1 - [\frac{(3-g_1)}{k_{n1}g_1(\mu_1 + \mu_2)}]^2 \tag{6a}$$

The steady state value of the state variable $x$ is $x_1^* = sin^{-1}(\frac{z_1^*}{g_1})$. The argument of arcsin is less than or equal to one in magnitude, therefore we get the upper limit of $d$ for AD state as given by,

$$[\frac{\beta_1\mu_1 - \beta_2\mu_2}{(\mu_1^2 - \mu_2^2)}] \leq g_1 \tag{6b}$$

Now, when $g_1 = g_2$ and $\beta_1 = \beta_2$, the final conditions of lower limit of $d$ is,

$$d > \left(\frac{1}{1-m}\right)[1 - \sqrt{\frac{\Omega_{n1}^2 + (3-g_1)^2}{(k_{n1}g_1)^2}}] \tag{7a}$$

And the upper limit of $d$ is,

$$d \leq \left(\frac{1}{1-m}\right)[1 - \frac{\Omega_{n1}}{k_{n1}g_1}] \tag{7b}$$

### b. Assessment of OD:

In case of OD, when the IHSS emerges through a symmetry breaking we have $x_1^* = -x_2^*$, $y_1^* = -y_2^*$, $z_1^* = -z_2^*$. Observing the relations given in equations (3a) and (3b), it is cleared that the OD through a symmetry breaking is obtained only when $g_1 = g_2$ and $\beta_1 = -\beta_2$.
To find the stability of the IHSS, we put $g_1 = g_2$ and $\beta_1 = -\beta_2$ in equations (6a) and (6b). Here, we get the conditions of lower limit of $d$ as follows,

$$[\Omega_{n1}\left(\frac{\mu_1 + \mu_2}{\mu_1 - \mu_2}\right)]^2 - [k_{n1}g_1(\mu_1 + \mu_2)]^2 > -(3 - g_1)^2 \tag{8a}$$

And the upper limit of $d$ is,

$$d \leq [1 - \frac{\Omega_{n1}}{k_{n1}g_1}] \tag{8b}$$

Note that, condition (8b) is independent of $m$.

### c. Assessment of transition between OD and AD:

Depending upon the coupling parameter if we want to make a transition from OD to AD, we have, $z_1^* = z_2^*$ with $\beta_1 \neq \beta_2$. This gives,

$$\mu_1 + \mu_2 = 0 \tag{9}$$

The equations (6a-6b) become undetermined on application of the above relation. Thus transition to AD is not possible in this process.

Another way to achieve the transition to AD is the way of frequency mismatch [1]. The characteristic frequency of the PLL depends on the time constant parameter $T(=RC)$ of the loop filter. Now, if we consider the two PLLs one of filter constant $T$ and another of $\alpha T$ (where $\alpha$ measures the frequency mismatch between the two PLLs), then the system equations are as follows,

$$\frac{dx_1}{d\tau} = \Omega_{n1} - k_{n1}(z_1 + d(m\bar{z} - z_1)) \tag{10a}$$

$$\frac{dy_1}{d\tau} = \sin(x_1) + (g_1 - 2)y_1 - (\frac{g_1-1}{g_1})z_1 \tag{10b}$$

$$\frac{dz_1}{d\tau} = g_1 y_1 - z_1 \tag{10c}$$

$$\frac{dx_2}{d\tau} = \alpha\Omega_{n2} - \alpha k_{n2}(z_2 + d(m\bar{z} - z_2)) \tag{10d}$$

$$\frac{dy_2}{d\tau} = \sin(x_2) + (g_2 - 2)y_2 - (\frac{g_2-1}{g_2})z_2 \tag{10e}$$

$$\frac{dz_2}{d\tau} = g_2 y_2 - z_2 \tag{10f}$$

But, the steady state value of the state variables $x$, $y$ and $z$ calculated from equation set (10) remains the same as before and does not contain the frequency mismatch parameter $\alpha$. So, it is not possible to achieve the transition from OD to AD through frequency mismatch.

We also use XPPAUT package [28] to compute the bifurcation branches. Fig- 2 (a) shows the bifurcation diagram of $x_1$ and $x_2$ with d for m = 0.2 and $g_1 = g_2 = 2.3$, $k_{n1} = k_{n2} = 0.4$ and $\Omega_{n1} = \Omega_{n2} = 0.3$. Bifurcation diagram has been drawn only about x=0 line. Bifurcation points about the other lines have been omitted for clear identification of points, where AD and OD starts. It is observed that at d = 0.24, AD is born through an inverse Hopf bifurcation. This AD (stable HSS) state becomes unstable trough a pitchfork bifurcation at d = 0.84. Fig- 2(b) shows the scenario for $g_1 = 2.2$, $g_2 = 2.3$, $k_{n1} = 0.6$, $k_{n2} = 0.4$, $\Omega_{n1} = 0.3$ and $\Omega_{n2} = -0.3$. Here, for d>0.3, the system enters into the OD state through pitchfork bifurcation. This quite supports our analytical predictions.

### 4. Numerical Simulation Results:

The dynamics of coupled TOPLL can be studied by examining the time domain evolution and the phase-plane plot of state variables. These are obtained through numerical integration of the state equations as given in equation (2). Note that equation set (2) and equation set (10) are identical if we put $\alpha = 1$. We use the 4th order Runge-Kutta technique of integration for this purpose. Every time, a large amount of initial data (about 80%) is discarded to eliminate the initial transients. The influences of $d$, $m$ and the conventional design parameters of PLL like $k_n$, $g$ etc. in the coupled PLL are obtained. It is observed that the self-oscillatory conditions of oscillations of isolated PLL may be period-1, period-2, period-4 etc types for different values of loop parameters and finally the dynamics of the system becomes chaotic.

At first, we examine the dynamics of the coupled system for two identical PLLs i.e. $g_1 = g_2$ and $\beta_1 = \beta_2$. We set $g_1 = g_2 = 2.3$, $k_{n1} = k_{n2} = 0.4$ and $\Omega_{n1} = \Omega_{n2} = 0.3$ for which each isolated PLL shows period-1 oscillation. Now we simulate the state equation (2) for different values of $d$ and $m$. We observe for proper choice of coupling strength and mean-field strength both the PLL of the coupled system converges to a HSS i.e. shows transitions from period-1 to AD. The results are shown through bifurcation diagram along with the numerical time series plots of $z_1$ and $z_2$ in Fig-3. We also simulate the region of AD in $(m - d)$ space for above mentioned design parameter values. Simulation results show that beyond the steady state, the coupled system exhibits limit cycle oscillations of period-1as well as multi period oscillations. The steady state (AD state) and the unstable oscillatory states of different periods are shown using different color sheds in Fig-4. The boundaries of the parameters indicating the AD state indicated by equations (7a) and (7b) are also plotted in the same figure. The simulation results agree very well with the analytically predicted results.

Next we consider the coupled system having the parameter values $g_1 = g_2$ and $\beta_1 = -\beta_2$. We set $g_1 = g_2 = 2.3$, $k_{n1} = k_{n2} = 0.4$, $\Omega_{n1} = 0.3$ and $\Omega_{n2} = -0.3$. In this case, we observe that for proper choice of $d$ and $m$ the coupled system shows transitions from period-1 to IHSS i.e. transition to OD state. The results are shown through bifurcation diagram along with the numerical time series plots of $z_1$ and $z_2$ in Fig-5. The OD state and the unstable oscillatory states of different periods are shown using different color sheds in Fig-5 for this case. The boundaries of the parameters indicating the OD state indicated by equations (8a) and (8b) are also plotted in the same figure. The simulation results agree very well with the analytically predicted results.

We also examine the dynamics of the coupled system having $g_1 = g_2 = 2.7$, $k_{n1} = k_{m2} = 0.4$, $\Omega_{n1} = 0.3$ and $\Omega_{n2} = -0.3$. With these values of design parameters each isolated PLLs are chaotic in nature. In this case also the coupled system converges to IHSS through inverse Hopf bifurcation with $d$ as the control parameter. We simulate the different dynamical states in $(m - d)$ space. The simulation results are shown in Fig-7. It indicates that the OD state reduces remarkably when PLLs are chaotic in nature. Here also the boundaries of the OD state are well calculated analytically by equations (8a) and (8b).

Finally, we extend our observation and examine the dynamics of the coupled system for two non-identical PLLs i.e. $g_1 \neq g_2$ and $\beta_1 \neq \beta_2$. We set $g_1 = 2.4$, $g_2 = 2.3$, $k_{n1} = 0.6$, $k_{n2} = 0.4$, $\Omega_{n1} = 0.1$ and $\Omega_{n2} = -0.3$ for which each isolated PLL shows period-1 oscillation. In this case we observe the transitions of period-1 to OD state, but the steady states are asymmetric in nature. When we plot the numerically simulated IHSS (i.e. OD state) and the unstable oscillatory states of different periods using different color sheds in Fig-8, the results shows reduce OD state. In this case, equations (6a) and (6b) are not able to calculate the boundaries of the OD state. Because the method we adopt here is applicable only when the transition occurs through a symmetry breaking. It has also been noted that in all cases the coupled system converges to trivial steady state values. We have not observed any transition between AD and OD in our simulation.

5. **Conclusion :**

We have explored the phenomena of oscillation quenching in the coupled TOPLL under the MFD coupling. The MFD coupling can induce AD and OD in the coupled system. But no transition between these two states has been observed. We have derived analytically the lower and upper limit of the oscillation quenched state using Routh-Hurwitz technique and predict the zone of AD/OD in the parameter space. The ability of convergence to quenched state of coupled PLLs depends on the design parameters. For identical system both the system converges to HSS state whereas for non-identical parameter values they converge to IHSS. It is also observed that for identical systems the quenched state is wider than non-identical case. When the system parameters are so chosen that each isolated PLLs are chaotic in nature, in that case we observe the quenched state is relatively narrow. All these numerical simulation results regarding system dynamics obtained applying nonlinear tools like bifurcation diagram are closely supported by the analytical predictions. The present study may be helpful in laser, neuronal and other biological systems and also in communication systems where phase locking phenomena takes place.

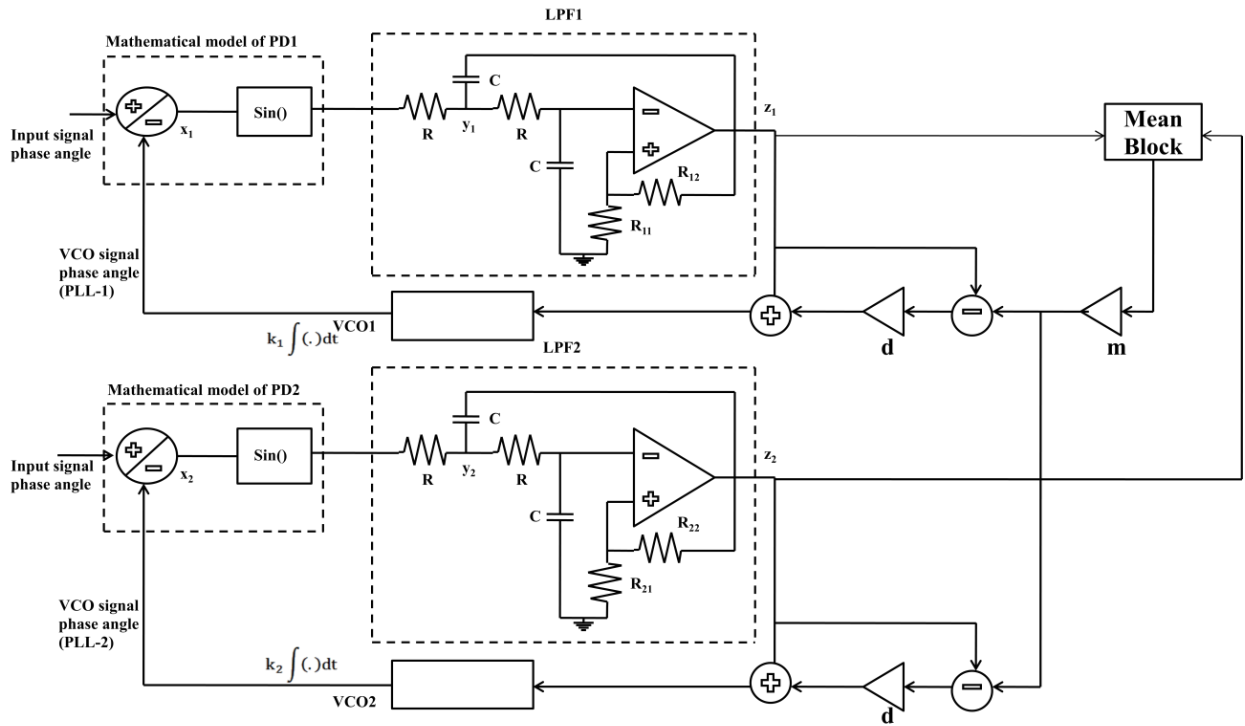

Fig-1: Block diagram of a coupled TOPLL with resonant low pass filter under MFD coupling.

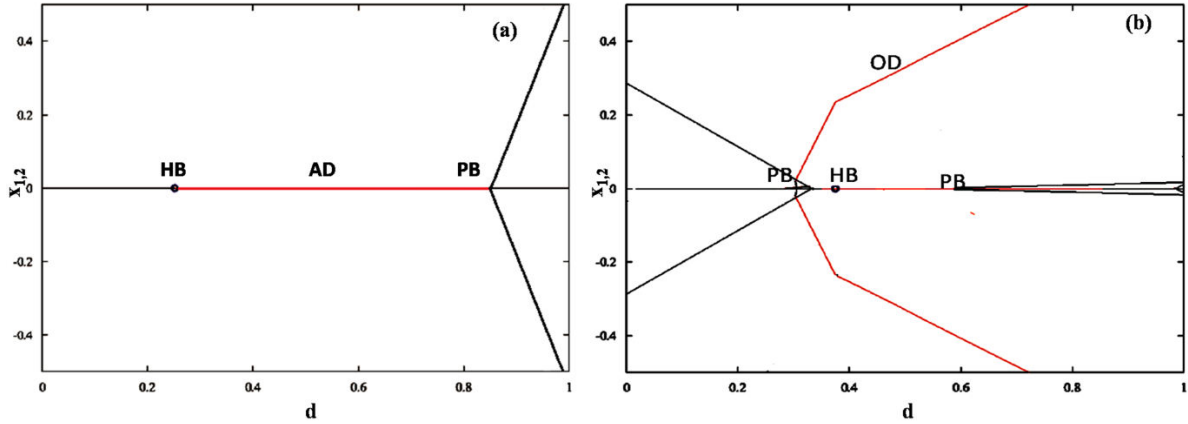

Fig.2 Bifurcation diagram (using XPPAUT) of two mean-field coupled TOPLLs. Red lines: stable fixed points, black lines: unstable fixed points. HB is Hopf bifurcation point, PB is pitchfork bifurcation points. (a) Dynamics of two identical PLL with m = 0.2 and $g_1 = g_2 = 2.3$, $k_{n1} = k_{n2} = 0.4$ and $\Omega_{n1} = \Omega_{n2} = 0.3$: AD takes place. (b) Dynamics of two non-identical TOPLLs with $g_1 = g_2 = 2.3$, $k_{n1} = k_{n2} = 0.4$, $\Omega_{n1} = 0.3$ and $\Omega_{n2} = -0.3$: OD takes place.

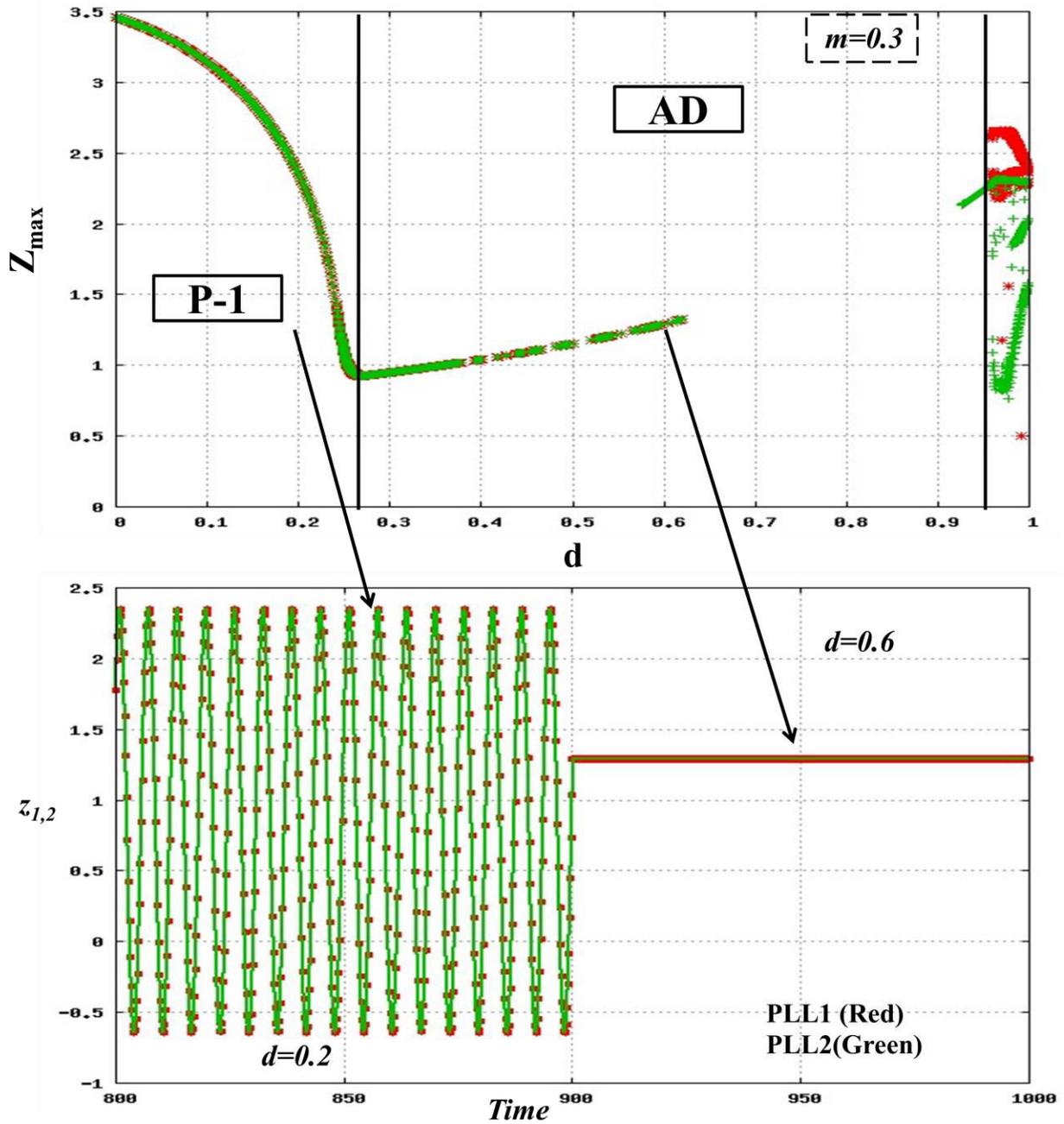

Fig-3: Numerically obtained bifurcation diagram (top) and time series plot (down) of $z_1$ and $z_2$ with $d$ as the control parameter. The values of other parameters are $g_1 = g_2 = 2.3$, $k_{n1} = k_{n2} = 0.4$ and $\Omega_{n1} = \Omega_{n2} = 0.3$ .

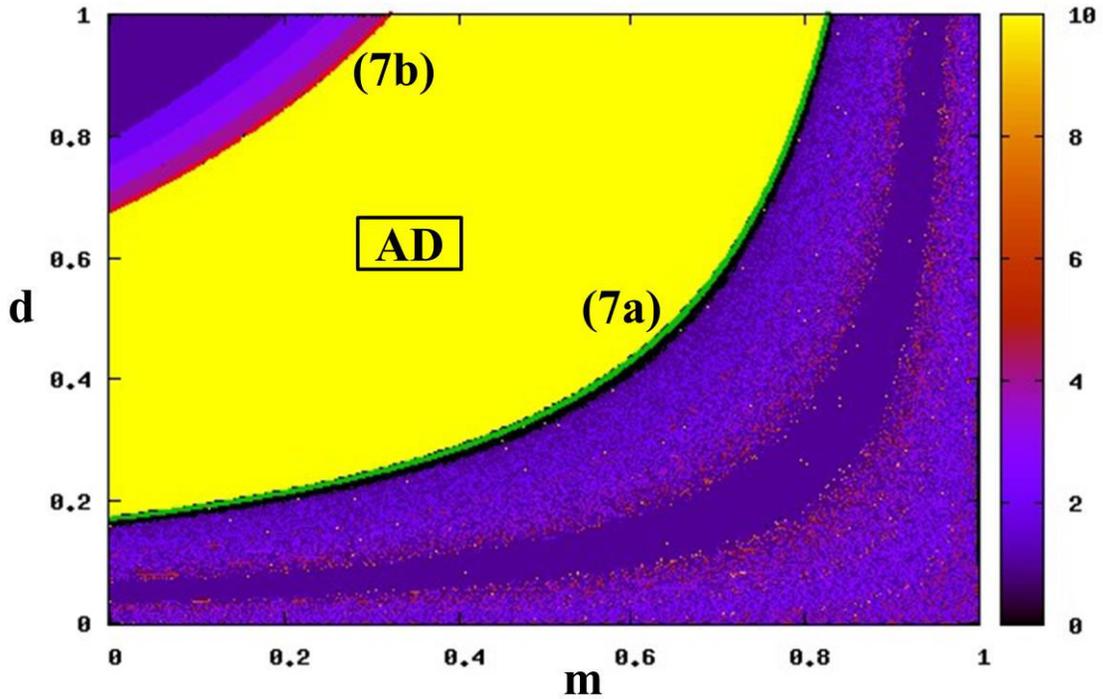

Fig-4: Various dynamical states of coupled PLL with $g_1 = g_2 = 2.3$, $k_{n1} = k_{n2} = 0.4$ and $\Omega_{n1} = \Omega_{n2} = 0.3$ shown in the $(m - d)$ parameter space, obtained by numerical simulation of state equations. The obtained states are, aperiodic state denoted by 0(black); oscillatory states of increasing period denoted by 1 to 9 (i.e. 1 for period-1, 2 for period-2 etc); AD state denoted by 10(green). The lower and upper bounds of the AD state as predicted analytically are shown by continuous lines.

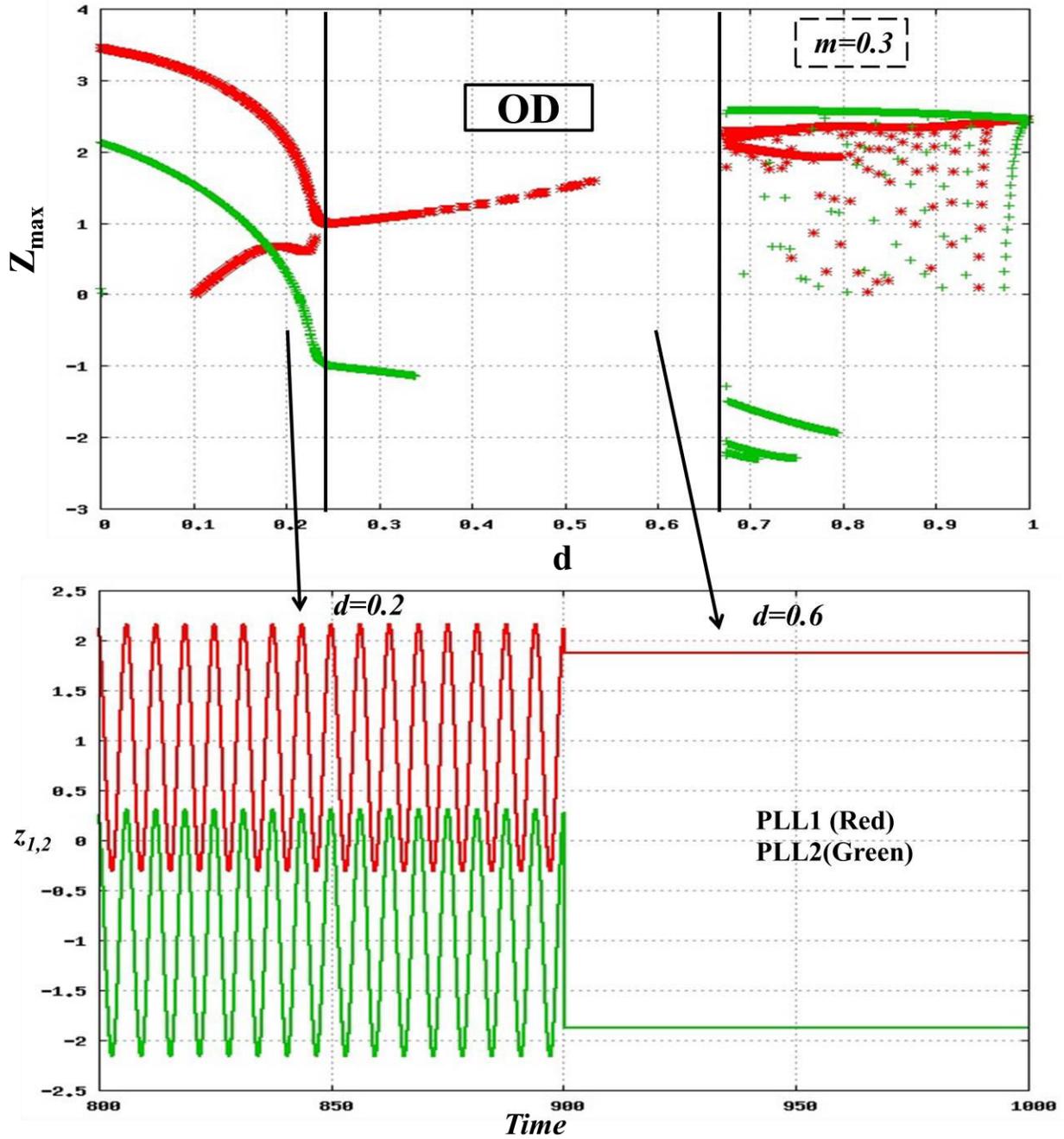

Fig-5: Numerically obtained bifurcation diagram (top) and time series plot (down) of $z_1$ and $z_2$ with $d$ as the control parameter. The values of other parameters are $g_1 = g_2 = 2.3$, $k_{n1} = k_{n2} = 0.4$ $\Omega_{n1} = 0.3$ and $\Omega_{n2} = -0.3$.

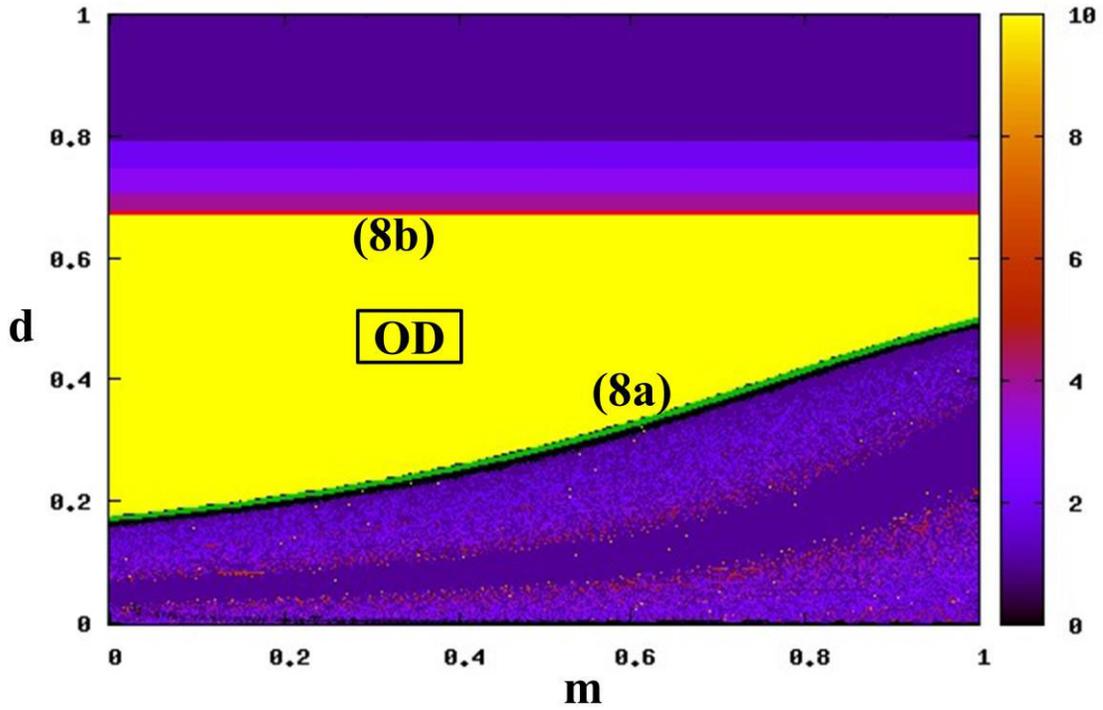

Fig-6: Various dynamical states of coupled PLL with $g_1 = g_2 = 2.3$, $k_{n1} = k_{n2} = 0.4$ $\Omega_{n1} = 0.3$ and $\Omega_{n2} = -0.3$ shown in the $(m - d)$ parameter space, obtained by numerical simulation of state equations. The obtained states are, aperiodic state denoted by 0(black); oscillatory states of increasing period denoted by 1 to 9 (i.e. 1 for period-1, 2 for period-2 etc); OD state denoted by 10(green). The lower and upper bounds of the OD state as predicted analytically are shown by continuous lines.

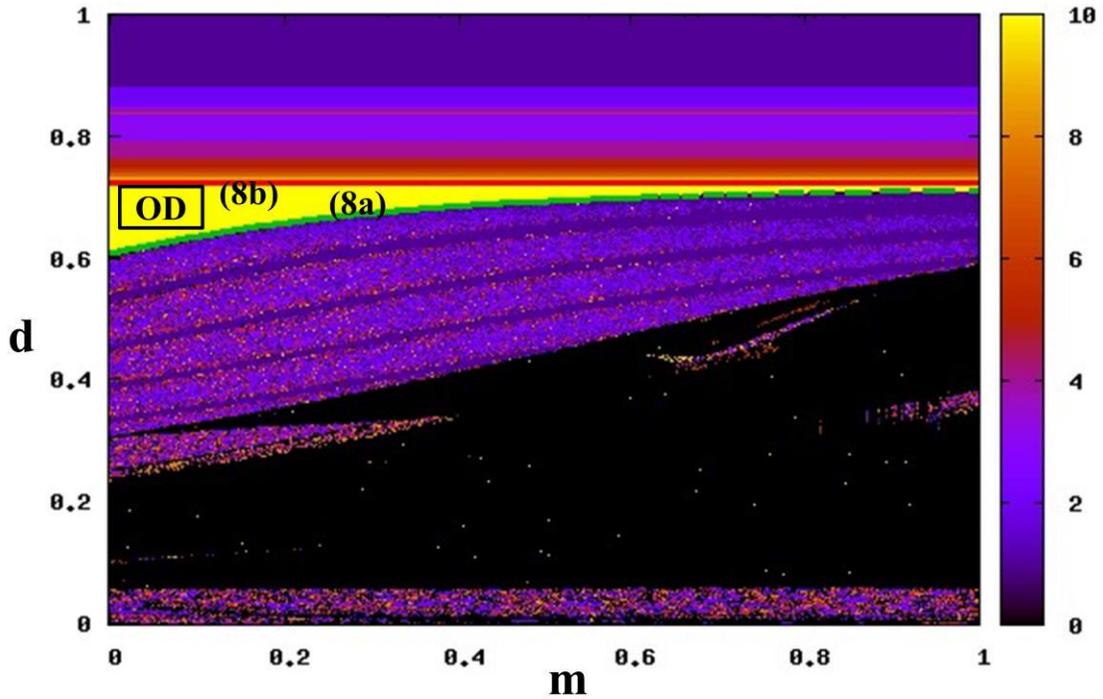

Fig-7: Various dynamical states of coupled PLL with $g_1 = g_2 = 2.7$, $k_{n1} = k_{n2} = 0.4$ $\Omega_{n1} = 0.3$ and $\Omega_{n2} = -0.3$ shown in the $(m - d)$ parameter space, obtained by numerical simulation of state equations. The obtained states are, aperiodic state denoted by 0(black); oscillatory states of increasing period denoted by 1 to 9 (i.e. 1 for period-1, 2 for period-2 etc); OD state denoted by 10(green). The lower and upper bounds of the OD state as predicted analytically are shown by continuous lines.

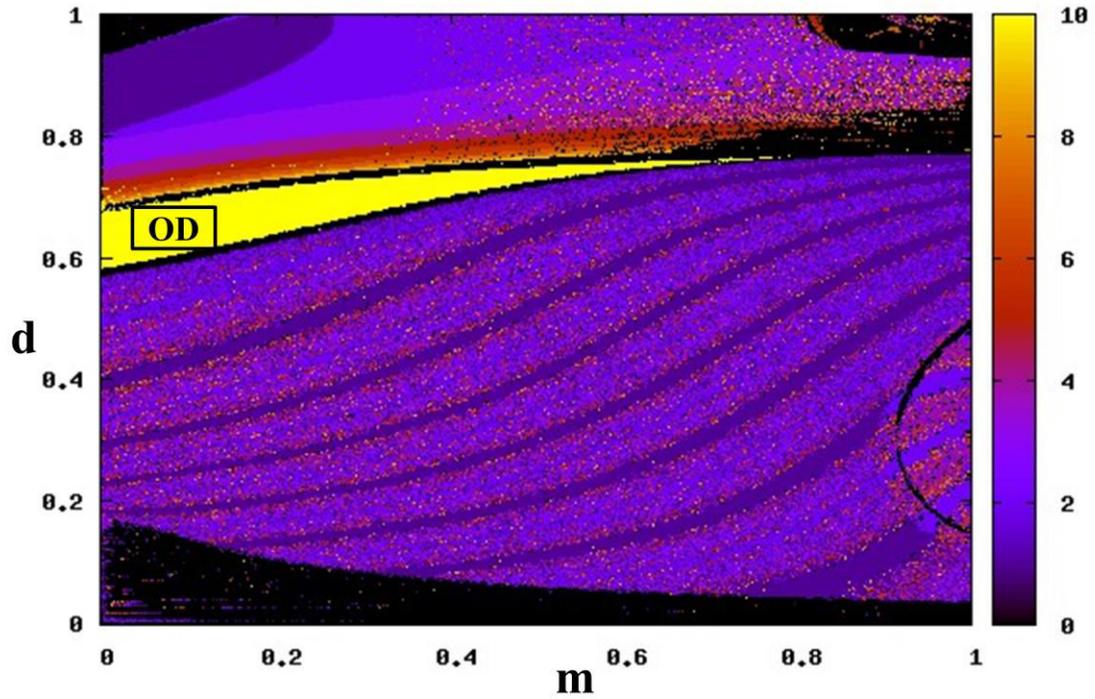

Fig-8: Various dynamical states of coupled PLL with $g_1 = 2.4$, $g_2 = 2.3$, $k_{n1} = 0.6$, $k_{n2} = 0.4$, $\Omega_{n1} = 0.1$ and $\Omega_{n2} = -0.3$ shown in the $(m-d)$ parameter space, obtained by numerical simulation of state equations. The obtained states are, aperiodic state denoted by 0(black); oscillatory states of increasing period denoted by 1 to 9 (i.e. 1 for period-1, 2 for period-2 etc); OD state denoted by 10(green).